\documentclass[a4paper,11pt]{article}
\usepackage{pos}
\usepackage{tikz}
\usetikzlibrary{calc}

\tikzset{%
  respieslice0/.style={fill=blue!40, draw=blue!72!black, line width=0.1pt},%
  respieslice1/.style={fill=violet!44, draw=violet!80!black, line width=0.1pt},%
  respieslice2/.style={fill=orange!46, draw=orange!82!black, line width=0.1pt},%
  respieslice3/.style={fill=teal!40, draw=teal!80!black, line width=0.1pt},%
  respieslice4/.style={fill=brown!42, draw=brown!78!black, line width=0.1pt},%
  resmarkbox0/.style={fill=blue!40, draw=blue!72!black, line width=0.34pt, rounded corners=1.15pt,
    minimum width=2.5mm, minimum height=2.5mm, inner sep=0pt},%
  resmarkbox0faint/.style={fill=blue!18, draw=blue!55!black, line width=0.24pt, rounded corners=1.15pt,
    minimum width=2.5mm, minimum height=2.5mm, inner sep=0pt},%
  resmarkbox1/.style={fill=violet!44, draw=violet!80!black, line width=0.34pt, rounded corners=1.15pt,
    minimum width=2.5mm, minimum height=2.5mm, inner sep=0pt},%
  resmarkbox1faint/.style={fill=violet!16, draw=violet!58!black, line width=0.24pt, rounded corners=1.15pt,
    minimum width=2.5mm, minimum height=2.5mm, inner sep=0pt},%
  resmarkbox2/.style={fill=orange!46, draw=orange!82!black, line width=0.34pt, rounded corners=1.15pt,
    minimum width=2.5mm, minimum height=2.5mm, inner sep=0pt},%
  resmarkbox2faint/.style={fill=orange!16, draw=orange!60!black, line width=0.24pt, rounded corners=1.15pt,
    minimum width=2.5mm, minimum height=2.5mm, inner sep=0pt},%
  resmarkbox3/.style={fill=teal!40, draw=teal!80!black, line width=0.34pt, rounded corners=1.15pt,
    minimum width=2.5mm, minimum height=2.5mm, inner sep=0pt},%
  resmarkbox3faint/.style={fill=teal!16, draw=teal!58!black, line width=0.24pt, rounded corners=1.15pt,
    minimum width=2.5mm, minimum height=2.5mm, inner sep=0pt},%
  resmarkbox4/.style={fill=brown!42, draw=brown!78!black, line width=0.34pt, rounded corners=1.15pt,
    minimum width=2.5mm, minimum height=2.5mm, inner sep=0pt},%
  resmarkbox4faint/.style={fill=brown!18, draw=brown!58!black, line width=0.24pt, rounded corners=1.15pt,
    minimum width=2.5mm, minimum height=2.5mm, inner sep=0pt},%
  resmarkboxfb/.style={fill=teal!40, draw=teal!80!black, line width=0.34pt, rounded corners=1.15pt,
    minimum width=2.5mm, minimum height=2.5mm, inner sep=0pt},%
  resmarkboxfbfaint/.style={fill=teal!16, draw=teal!58!black, line width=0.24pt, rounded corners=1.15pt,
    minimum width=2.5mm, minimum height=2.5mm, inner sep=0pt},%
}

\newcommand{\resmarkcross}[1]{%
  \node[#1] (rmb) {};%
  \draw[line width=0.82pt, line cap=round, black]%
    ($(rmb.center)+(-0.52mm,-0.52mm)$) -- ($(rmb.center)+(0.52mm,0.52mm)$)%
    ($(rmb.center)+(-0.52mm,0.52mm)$) -- ($(rmb.center)+(0.52mm,-0.52mm)$);%
}

\newcommand{\resmark}[2]{%
  \begin{tikzpicture}[baseline=-0.55ex]
    \ifcase\numexpr#1\relax
      \ifnum#2=0\relax\node[resmarkbox0] {};
      \else\resmarkcross{resmarkbox0}\fi
    \or
      \ifnum#2=0\relax\node[resmarkbox1] {};
      \else\resmarkcross{resmarkbox1}\fi
    \or
      \ifnum#2=0\relax\node[resmarkbox2] {};
      \else\resmarkcross{resmarkbox2}\fi
    \or
      \ifnum#2=0\relax\node[resmarkbox3] {};
      \else\resmarkcross{resmarkbox3}\fi
    \or
      \ifnum#2=0\relax\node[resmarkbox4] {};
      \else\resmarkcross{resmarkbox4}\fi
    \else
      \ifnum#2=0\relax\node[resmarkboxfb] {};
      \else\resmarkcross{resmarkboxfb}\fi
    \fi
  \end{tikzpicture}%
}

\newcommand{\ressublabel}[3]{#3}

\newsavebox{\proceedingsquarkbox}
\newcommand{\proceedingsQuarkFigWidth}{1.375cm}

\newcommand{\ProceedingsQuarkFigTwo}[2]{%
  \sbox{\proceedingsquarkbox}{%
    \includegraphics[width=\proceedingsQuarkFigWidth]{#1}\hspace{1.5pt}%
    \includegraphics[width=\proceedingsQuarkFigWidth]{#2}}%
  \raisebox{\dimexpr(\ht\strutbox-\dp\strutbox-\ht\proceedingsquarkbox+\dp\proceedingsquarkbox)/2\relax}{%
    \usebox{\proceedingsquarkbox}}%
}

\newcommand{\SubsectionQuarkTwo}[3]{%
  \subsection{\texorpdfstring{%
      \begin{minipage}[t]{0.68\linewidth}\raggedright #1\end{minipage}%
      \hfill
      \ProceedingsQuarkFigTwo{#2}{#3}%
    }{#1}}}

\graphicspath{{figures/quark_diagrams/}{figures/}}

\title{Exotic Hadron Spectroscopy in Heavy-Flavor Systems}
\ShortTitle{Exotic heavy-flavor spectroscopy}

\author*[a]{M. Mikhasenko}

\affiliation[a]{Ruhr University Bochum,\\
  Universitätsstraße 150, Bochum, Germany}

\emailAdd{mikhail.mikhasenko@cern.ch}

\abstract{%
  Recent years have brought a dense sequence of experimental discoveries in heavy-flavor hadron spectroscopy.
  Heavy-flavor spectroscopy has entered a period in which new hadronic structures are no longer isolated surprises but recurring features across several flavor sectors, seen by multiple experiments in several decay environments. For systems with charm and bottom quarks, smaller widths and cleaner signatures expose regularities that would be harder to isolate in the light-quark sector.
  This contribution focuses on the classes of states that now define the modern ``exotic'' landscape: hidden-charm pentaquarks,
  charged charmonium-like structures, resonances in onia-onia systems, doubly-heavy tetraquarks, and open-flavor tetraquarks.}

\FullConference{Proceedings of the 53rd International Symposium on Multiparticle Dynamics (ISMD 2025)\\
Corfu, Greece\\}

\begin{document}
\maketitle

\section{Introduction}

Quantum chromodynamics (QCD) is the theory of quarks and gluons, but the hadronic world it produces is not a weakly coupled parton picture.
The interaction is nonperturbative in that regime, color is confined, and the degrees of freedom that show up in experiments are hadrons and their excitations.
Perturbation theory in the fundamental fields becomes reliable at energies well above the QCD scale, $\Lambda_{\rm QCD}\sim 300\,\mathrm{MeV}$.
It marks the boundary below which the dynamics of light hadrons and soft processes are intrinsically nonperturbative.
The $u$, $d$, and $s$ quarks are light on hadron scales. They should not be described as localized objects.
At scales below $\Lambda_{\rm QCD}$ they are better viewed as delocalized color degrees of freedom, strongly mixed with gluons and sea-quark excitations.
Charm and bottom quarks are much heavier, and in heavy hadrons they can be treated as almost pointlike, slowly moving color sources.
Even for states with heavy quarks, most states of interest are observed as resonant enhancements in invariant-mass spectra rather than as directly reconstructed long-lived particles.
In the valence picture, the simplest color singlets are conventional mesons ($q\bar{q}$) and baryons ($qqq$).
Hadrons that require a larger multiquark valence content, such as tetraquark or pentaquark topologies, are classified as exotics.

Studies of these short-lived particles are performed statistically, by analyzing a large number of reactions.
Probability (the rate of occurrence) is used as a proxy for the underlying particle dynamics, which cannot be accessed by registering individual particles.
Here is an intuitive way to understand the central idea behind the statistical analysis of excited hadron states as resonances.
Consider a scattering setup with elastic reactions $AB\to AB$:
(a)~we set the total energy of the system from the momentum of the colliding particles;
(b)~we bring them into collision;
(c)~we measure the interaction probability, repeating the collision many times;
(d)~we then change the energy and repeat the procedure to obtain the probability at neighboring energies.
The probability-density profile measured in these studies represents a physical quantity related to the cross section. Peaks in this profile can be interpreted as signatures of resonances.

\begin{figure}[h]
  \centering
  \includegraphics[width=0.5\textwidth]{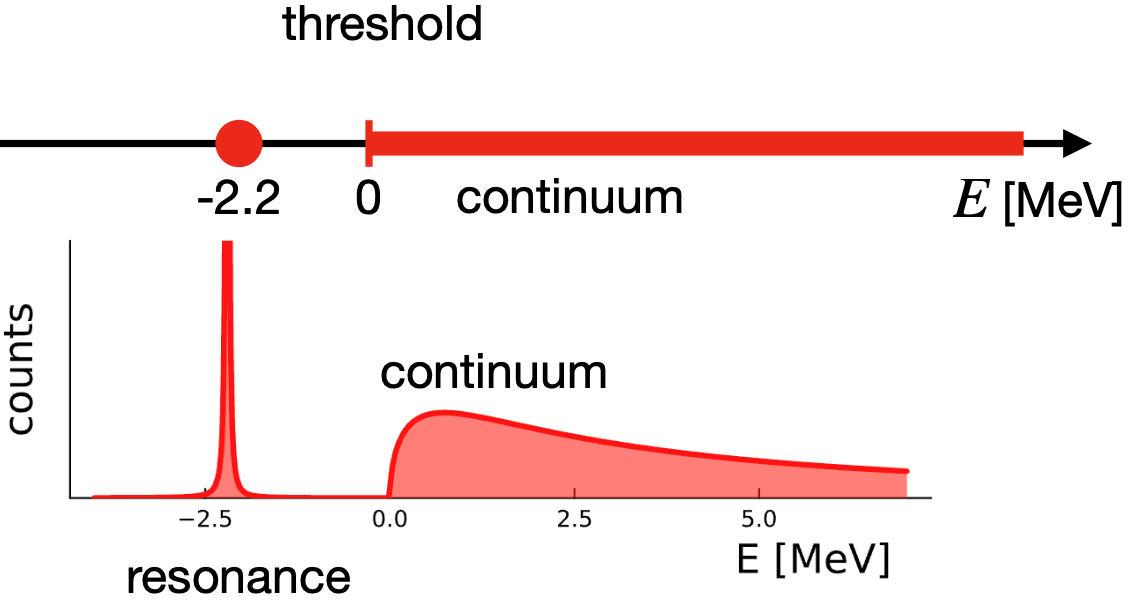}
  \caption{A sketch of a deuteron-like resonance seen from a weakly coupled inelastic channel.}
  \label{fig:deuteron-resonance}
\end{figure}

At the same time, not all interesting states should be thought of as simple isolated Breit--Wigner peaks.
The deuteron remains the textbook example of a hadronic molecule.
It is bound below threshold in the elastic channel of the proton--neutron system, and it does not appear as a resonance peak in that system.
The deuteron is stable, but if there were weak coupling to an inelastic channel with a lower threshold, one could observe an extremely narrow peak associated with the deuteron, and possibly a threshold enhancement related to the $pn$ channel opening as shown in Fig.~\ref{fig:deuteron-resonance}.
Much of the current discussion of heavy exotic states uses this analogy.
Many observed narrow states, particularly with charm--(anti)charm quarks, manifest themselves as hadronic molecular structures,
anchored to a threshold.

Hadrons and their excitations are studied in two main ways:
exclusive decays in which the states appear as intermediate states in cascade decays, e.g.\ in b-hadron decays (Belle~II, LHC),
or in direct production, such as $e^+e^-$ collisions (BESIII), or inclusive prompt processes in $pp$ collisions (LHC).

The purpose of this contribution is not to give a complete review,
but to highlight the recent measurements that most clearly shape the present experimental landscape.
This contribution is a personal selection of reactions and states, and the emphasis is necessarily biased by the author's work in the LHCb collaboration.
The material is organized primarily around experimental spectra, with only a limited number of theoretical references made explicit.
For structures seen in more than one channel, experimental spectra are arranged with aligned mass axes so that the same physical content can be compared side by side.
Several figures are adapted from the original publications in order to stress features that matter for the discussion.
The $B \to D\bar{D}h$ discussion in Section~\ref{sec:B2DDh} is developed by systematically tabulating the available charge- and flavor-combination topologies, so that the combinatorial structure of the measurement program becomes transparent.

\SubsectionQuarkTwo{Pentaquarks}{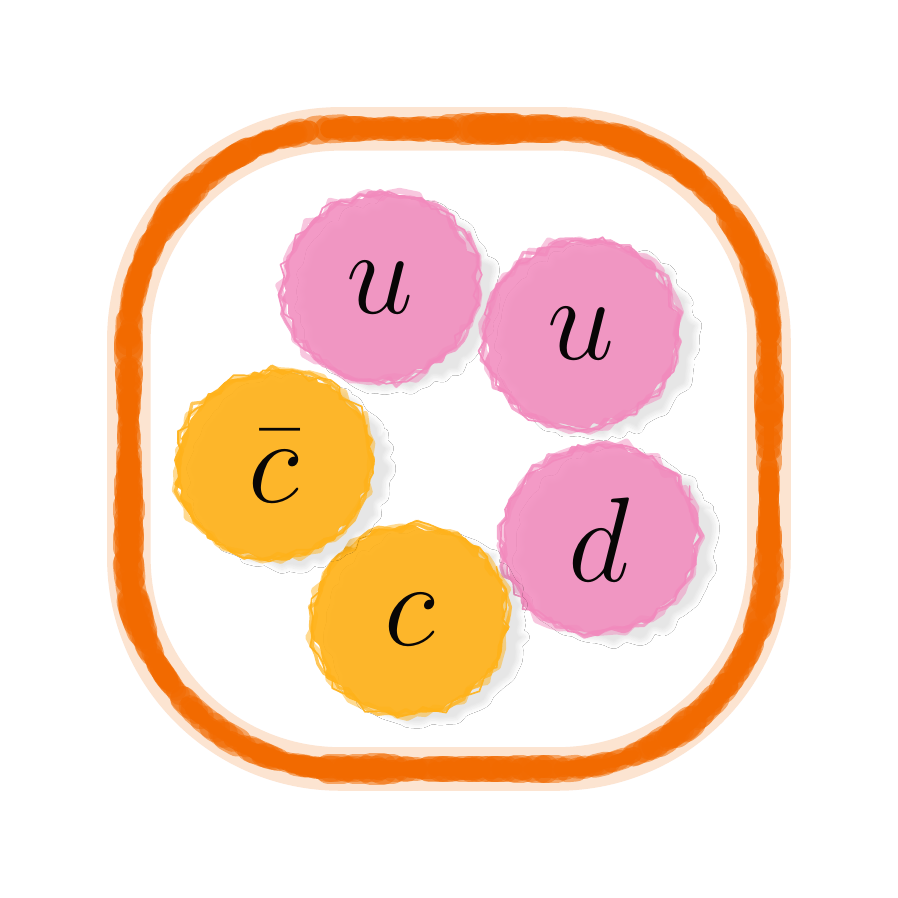}{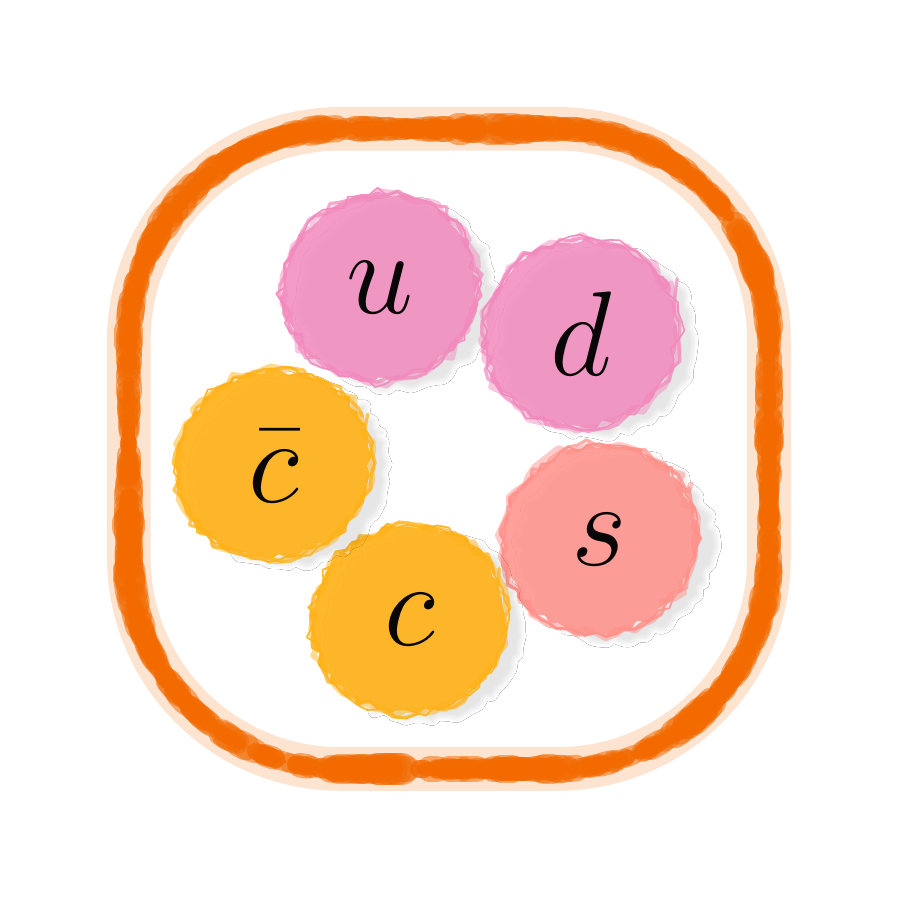}

Experimental evidence for hidden-charm pentaquark candidates has emerged from $b$-decay studies in the LHCb experiment.
The 2015 amplitude analysis of the $\Lambda_b^0 \to J/\psi K^- p$ decays revealed a resonant structure in the $J/\psi p$ channel consistent with hidden-charm pentaquark states~\cite{LHCb:2015yax,LHCb:2016ztz}.
A higher-statistics update in 2019 refined this picture.
The earlier prominent peaking structure was resolved into two narrower components, $P_{c\bar{c}}(4457)^+$ and $P_{c\bar{c}}(4440)^+$~\cite{LHCb:2019kea}.
Another narrow state, $P_{c\bar{c}}(4312)^+$, was observed in the same spectrum.
The broad state $P_{c\bar{c}}(4380)^+$, introduced in the 2015 analysis as an opposite-parity counterpart to the structure around $4450\,\mathrm{MeV}$, was not needed to describe the $m(J/\psi p)$ spectrum.
Admittedly, the sensitivity of the one-dimensional analysis was limited to prominent narrow features.
Several points in the angular-analysis formalism have been scrutinized in the literature~\cite{JPAC:2019ufm}, which enabled an accurate cross-check of the construction of the hadronic amplitude~\cite{Wang:2020giv,Habermann:2024sxs}.
The updated multi-dimensional analysis is work in progress.

A further extension of the hidden-charm pentaquark family came with the 2023 observation of a $J/\psi \Lambda$ resonance in $B^- \to J/\psi \Lambda \bar p$, providing evidence for a strange analogue of the hidden-charm pentaquark signal~\cite{LHCb:2022ogu}.

\begin{figure}[h]
  \centering
  \includegraphics[width=0.9\textwidth, trim=0 30mm 0 50mm, clip]{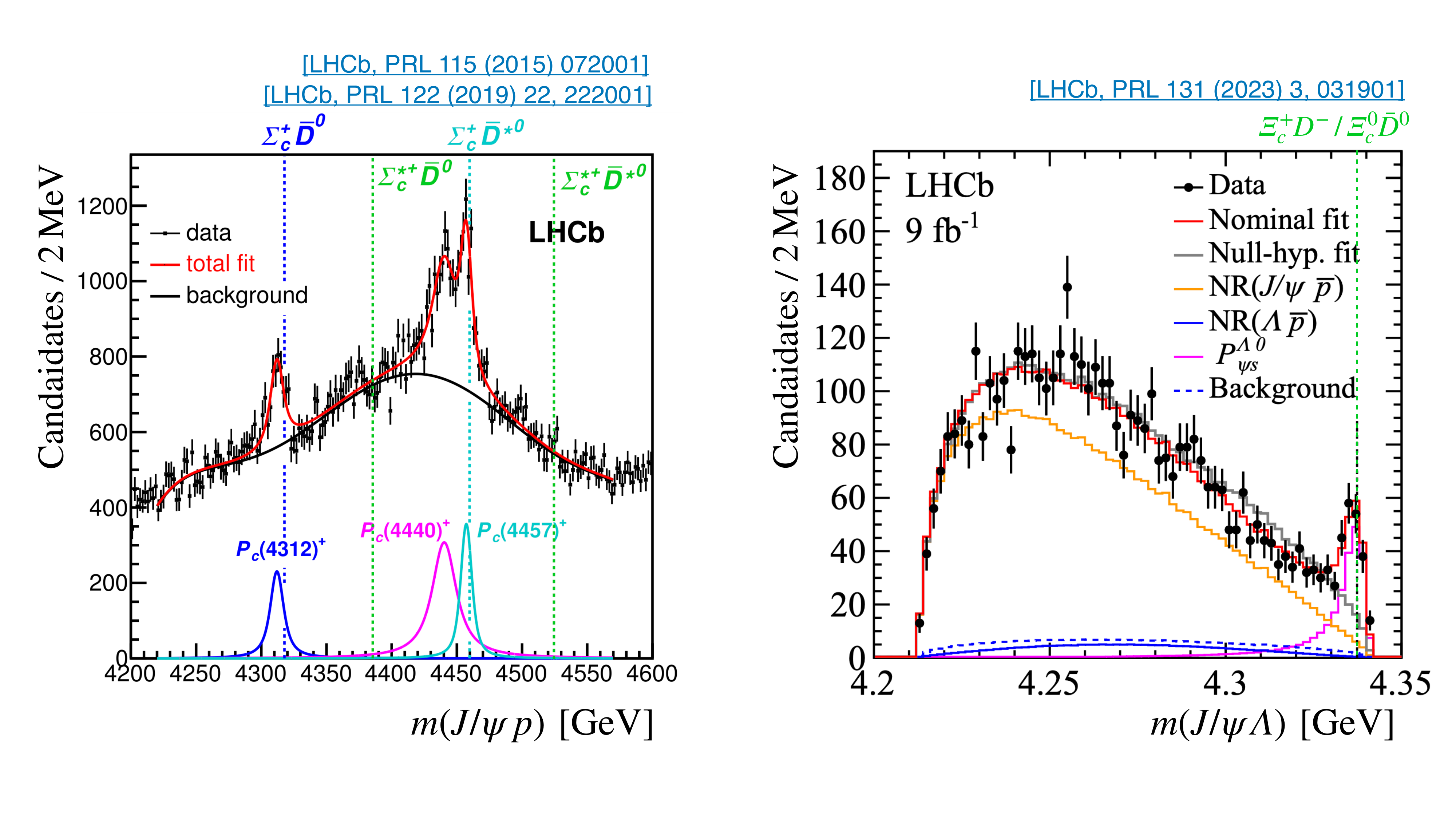}
  \caption{Hidden-charm pentaquark candidates in the $J/\psi p$ channel~\cite{LHCb:2015yax} and the $J/\psi \Lambda$ channel~\cite{LHCb:2022ogu}.}
  \label{fig:hidden-charm-pentaquark}
\end{figure}

The central point in the present understanding of the narrow pentaquark states is that they lie very close to the meson-baryon thresholds.
For the $J/\psi p$ system, relevant thresholds are $\Sigma_c^{(*)} \overline{D}^{(*)}$.
For the $J/\psi\,\Lambda$ case, these are thresholds of the $\Xi_c\overline{D}$ system.
This proximity suggests an important role for continuum, meson-baryon dynamics in the appearance of the pentaquark states.
A nearby threshold alone does not prove the internal structure of a state.
The present data remain compatible with several interpretations.
It is worth noting that the SU(3) multiplet partner of $\Sigma_c^+$ is the $\Xi_c^{\prime0}$ baryon, not the $\Xi_c^0$ baryon.

Further hints of pentaquark states come from the $B_s^0 \to p\bar{p} J/\psi$ decay~\cite{LHCb:2021chn} and $\Xi_b^- \to J/\psi \Lambda K^-$ decay~\cite{LHCb:2020jpq}.
Several prospective reactions with charmonium-baryon components~\cite{%
  LHCb:2025lhk,
  LHCb:2020kkc,
  LHCb:2019imv
}, and open-flavor meson-baryon systems~\cite{%
  LHCb:2025lwm,
  LHCb:2023eeb,
  LHCb:2024fel,
  LHCb:2024pnt
} have been observed, albeit limited by statistics.
The LHC Run~3 data, currently being collected and analyzed, may provide new evidence for pentaquark resonances.

\SubsectionQuarkTwo{Charged charmonium-like states}{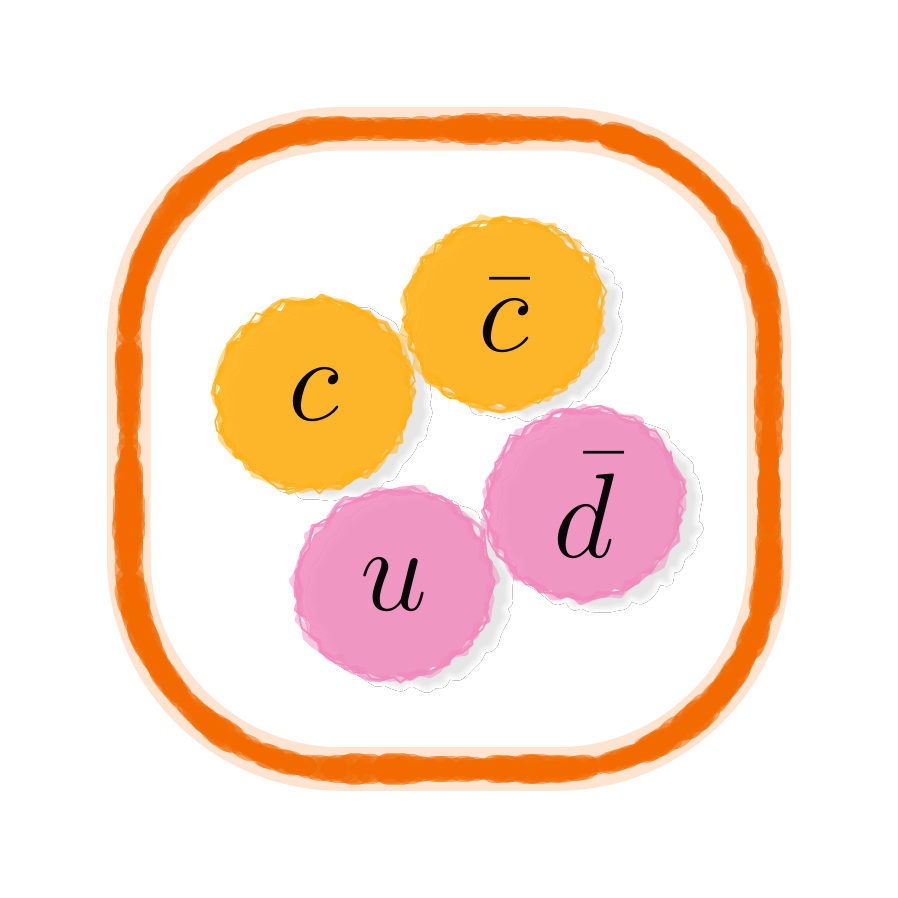}{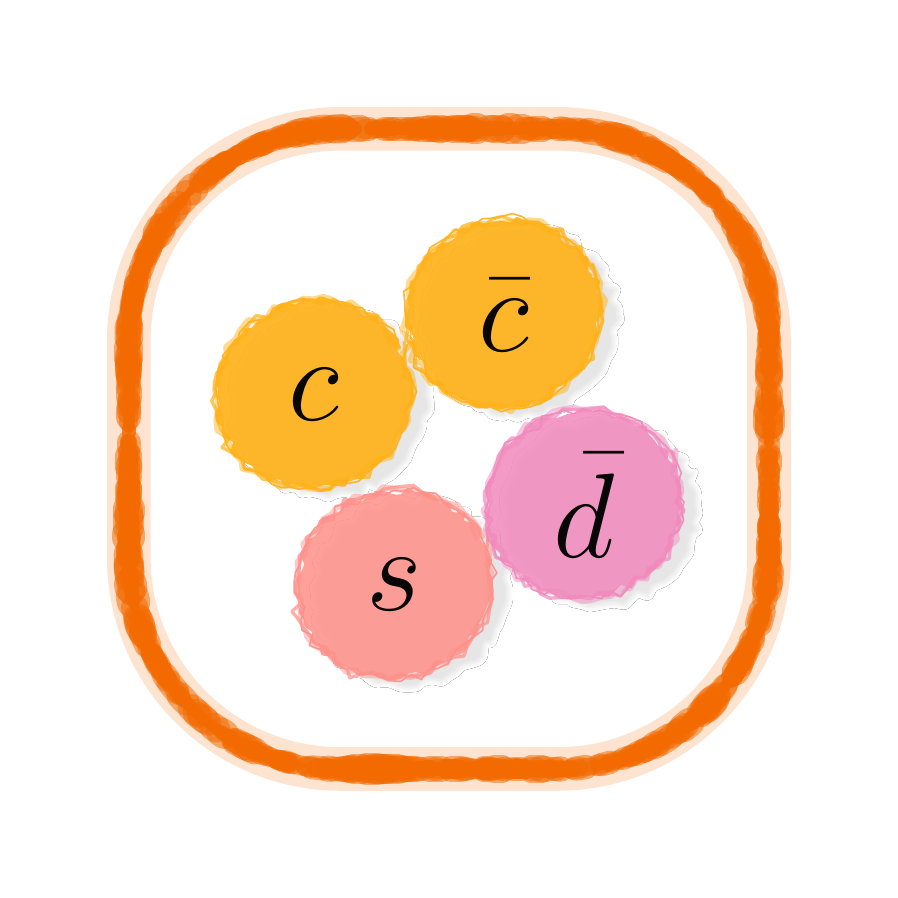}

\textit{Charged} charmonium-like candidates provide one of the clearest demonstrations that heavy-hadron spectroscopy extends beyond conventional quarkonium.
In current naming conventions~\cite{ParticleDataGroup:2024cfk} these states are written as $T_{c\bar{c}}^+$ and $T_{c\bar{c}s}^+$, while older papers and experimental titles frequently employ the original labels $Z_c$ and $Z_{cs}$.

The experimental story begins with the $T_{c\bar{c}}^+(4430)$ state reported by Belle as a resonance-like enhancement in the $\pi^+ \psi(2S)$ spectrum seen in $B \to K^+ \pi^- \psi(2S)$ decays~\cite{Belle:2007hrb}. Later, LHCb confirmed it and established its quantum numbers as $J^{P} = 1^{+}$~\cite{LHCb:2014zfx,LHCb:2015sqg}. The state also appears in
$B^+ \to \psi(2S) K^+ \pi^+ \pi^-$ decays with lower statistics~\cite{LHCb:2024cwp}. It is worth noting that statistics in studies of $B \to J/\psi K^+ \pi^-$ and $B \to \psi K^+ \pi^-$ are no longer the limiting factor, and the challenge lies in the modeling domain~\cite{Beiter:2023ltc}.
A recent LHCb update in $B^+ \to \psi(2S) K_S^0 \pi^+$ presents an isospin-related channel suited to studying the $T_{c\bar{c}}^+$ structures~\cite{LHCb:2025kxf}.

The long-lived reference point for this sector is the charged charmonium-like tetraquark $T_{c\bar{c}}^+(3900)$, first seen by BESIII in $e^+e^- \to \pi^+\pi^- J/\psi$, where a pronounced structure appears in the $J/\psi \pi^+$ system~\cite{BESIII:2013ris}.
The state is seen in many channels in analyses of the BESIII, Belle, and CLEO-c data samples. Reference~\cite{BESIII:2026ret} presents a state-of-the-art combined analysis of additional structure near $4.02\,$GeV, together with a useful overview of the literature.

\begin{figure}[h]
  \centering
  \includegraphics[width=0.9\textwidth]{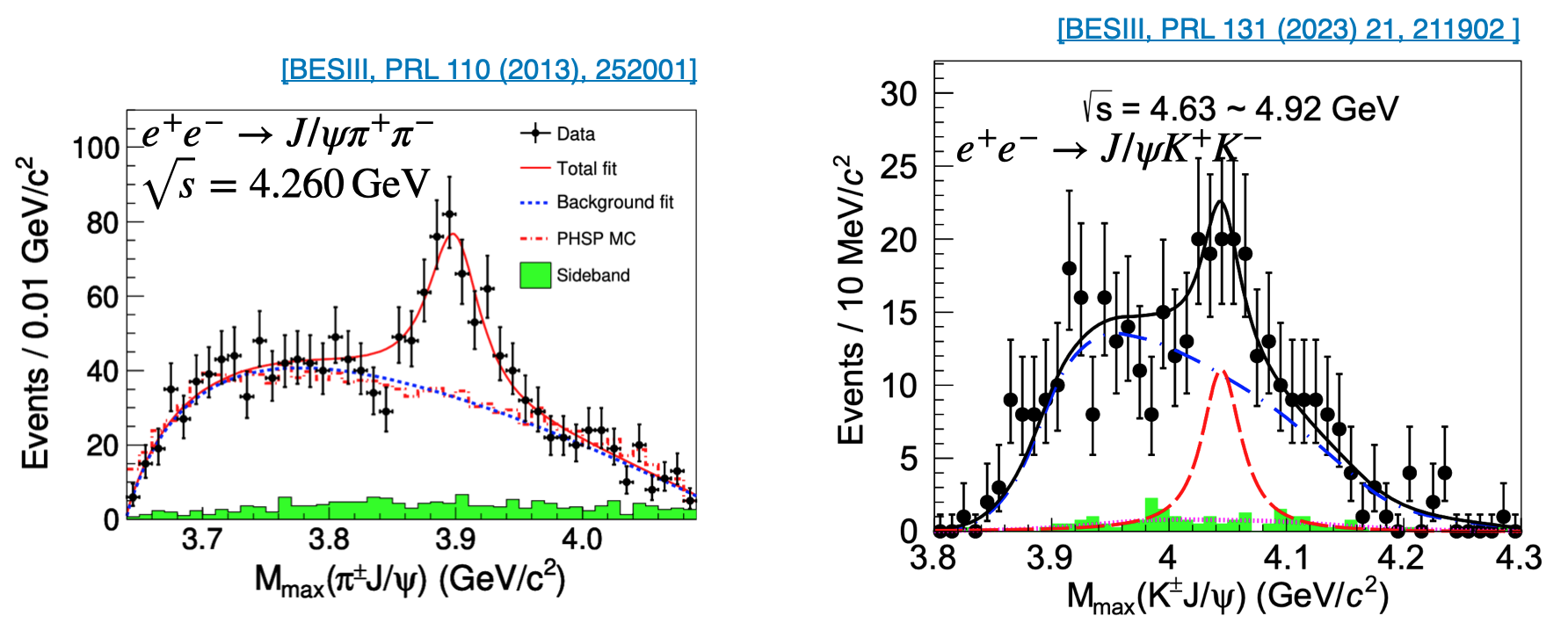}
  \caption{Charmonium-like structure in $e^+e^- \to \pi^+\pi^- J/\psi$ and $K^+K^- J/\psi$ final states from the BESIII experiment.}
  \label{fig:charged-charmonium-like-states}
\end{figure}

The BESIII experiment has also reported a vector charmonium-like structure near $4.7\,\mathrm{GeV}$ in the $J/\psi K^-$ system in the $e^+ e^- \to K^+K^- J/\psi$ process~\cite{BESIII:2023wqy}, as shown in Fig.~\ref{fig:charged-charmonium-like-states}. The state is discussed as a $T_{c\bar{c}s}^+$-type strange partner of $T_{c\bar{c}}^+(3900)$~\cite{Yang:2020nrt,Meng:2020ihj,Maiani:2021tri}. 

The charmonium-like sector is very complex, and many hadronic thresholds may be relevant for understanding the dynamics and for explaining the appearance of these states.
To identify relevant degrees of freedom and elucidate the interplay among these states, three directions can be listed:
(a)~observe these states in new decay modes and map their relative contributions;
(b)~observe new states and group them into spin multiplets;
(c)~observe new states in isospin- and SU(3)-related channels.
The amount of data already available, together with the growing sophistication of amplitude analyses, makes the outlook for this sector encouraging.

\SubsectionQuarkTwo{Onia-onia and fully-heavy structures}{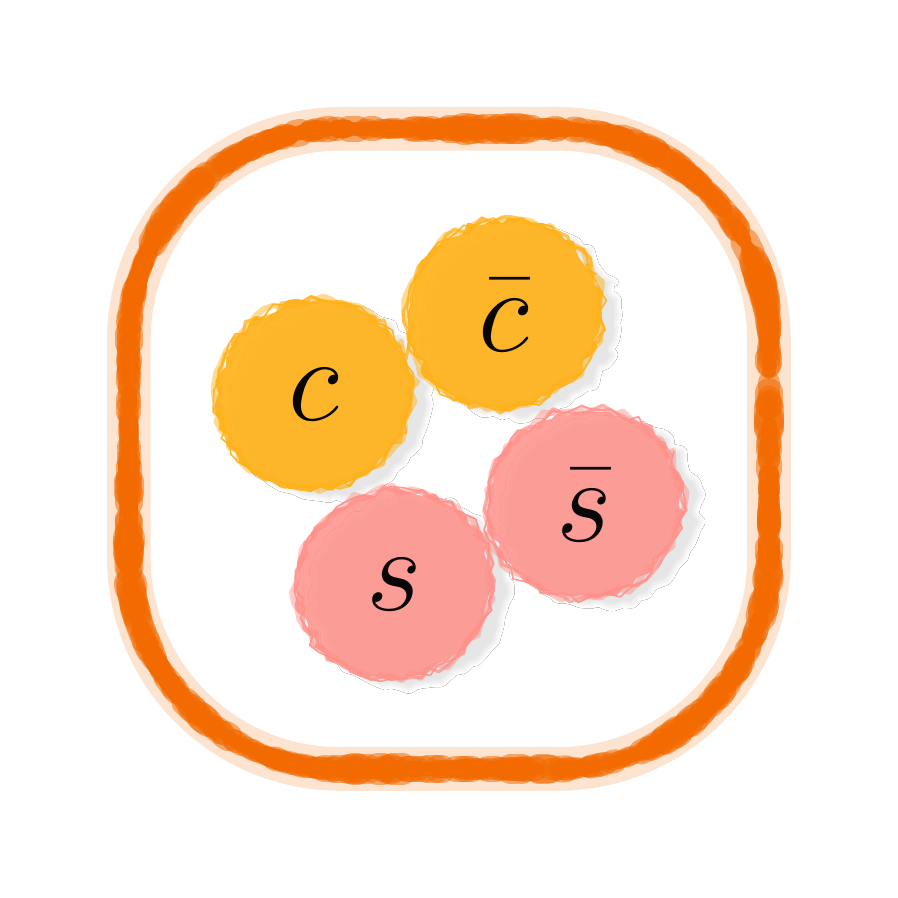}{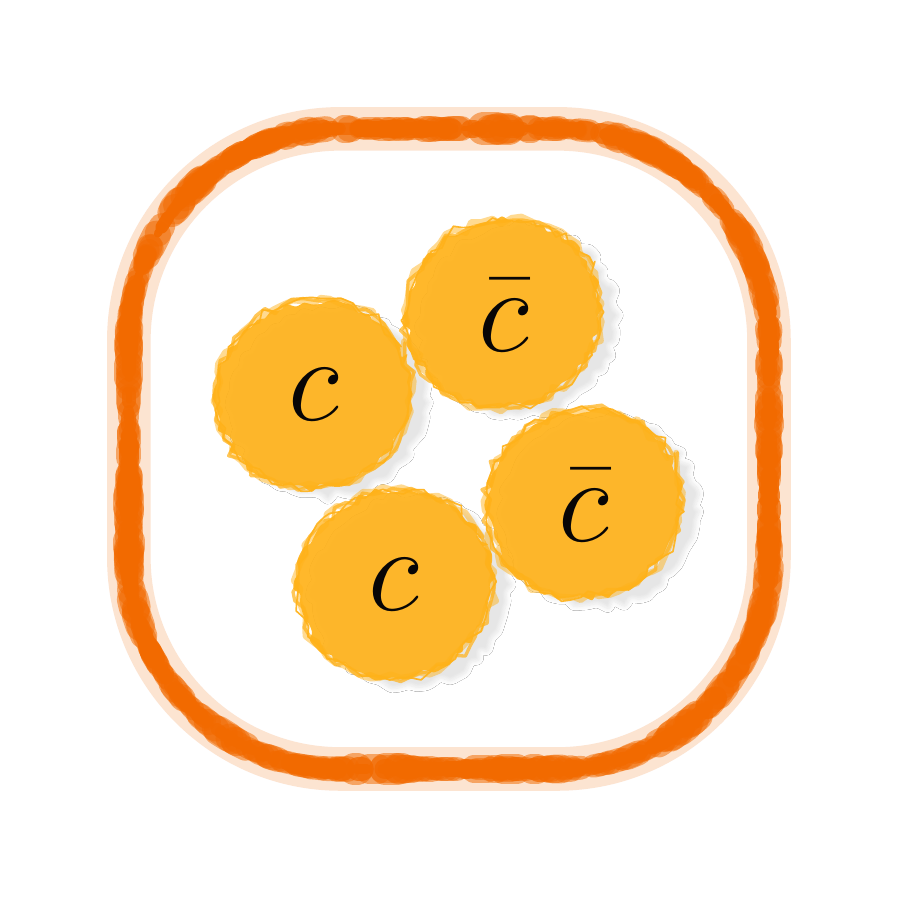}

Manifestly exotic states of the previous section must have neutral partners within the same isospin multiplet.
They are studied in decays to final states that are also coupled to conventional charmonia.
The latter already form a dense resonance spectrum, making it hard to distinguish regular from exotic states. Some spectra are particularly spectacular.

\begin{figure}[h]
  \centering
  \includegraphics[width=0.9\textwidth]{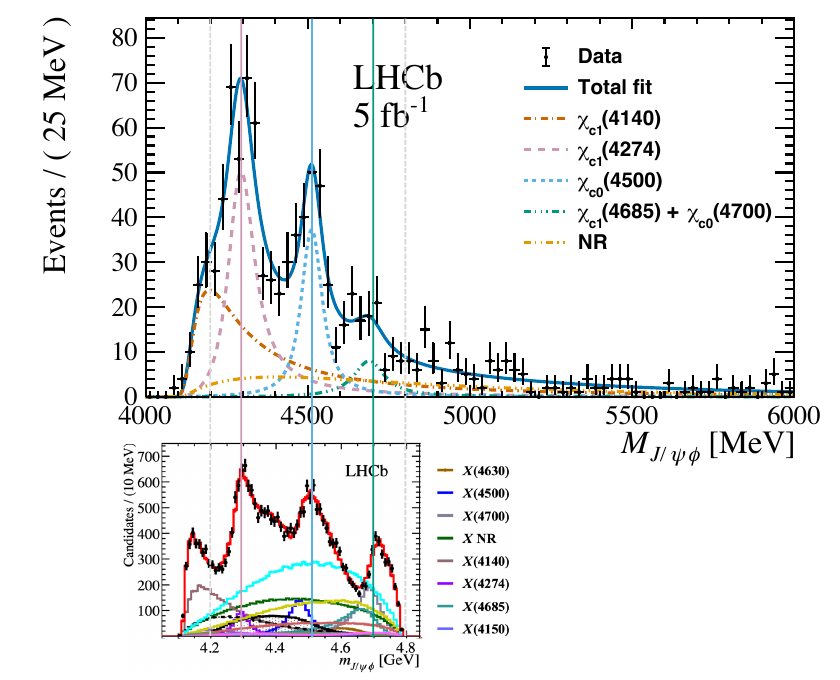}
  \caption{$J/\psi \phi$ resonances seen in $B$-decay modes $B\to J/\psi \phi K$~\cite{LHCb:2021uow} and in central exclusive production at LHCb~\cite{LHCb:2024smc}.}
  \label{fig:onia-onia-and-fully-heavy-structures}
\end{figure}

The spectrum of $J/\psi\,\phi$ seen in the $B^+ \to J/\psi\,\phi\,K^+$ decays
is densely populated with resonances~\cite{LHCb:2016axx}.
The latest LHCb amplitude analysis of this system~\cite{LHCb:2021uow}
employs a model that includes new $J/\psi\,K^+$ and $J/\psi\,\phi$ structures.
The same $J/\psi\,\phi$ system was also observed in a completely different production environment, central exclusive production in proton--proton collisions~\cite{LHCb:2024smc}.
Fig.~\ref{fig:onia-onia-and-fully-heavy-structures} shows that the two spectra display similar patterns.
Amplitude analyses favor positive parity, with spin assignments $J=0$ and $J=1$, while the identification of these states as exotic or conventional remains open.

Among the more surprising systems showing resonance structure are double-charmonium systems, in particular $J/\psi J/\psi$ and $J/\psi \psi(2S)$. The former was observed by the LHCb Collaboration~\cite{LHCb:2020dijpsi}, and also by ATLAS~\cite{ATLAS:2023bft} and CMS~\cite{CMS:2023owd,CMS:2025fpt}. The most recent update from CMS suggests three resonance structures in the di-$J/\psi$ spectrum that are broadly consistent with the features of the $J/\psi\,\psi(2S)$ spectrum.
For an overview of developments in this channel,
I refer to the talk and proceedings of Kai Yi~\cite{KaiYi.Corfu:2025xyz}.

\SubsectionQuarkTwo{Doubly-heavy tetraquarks}{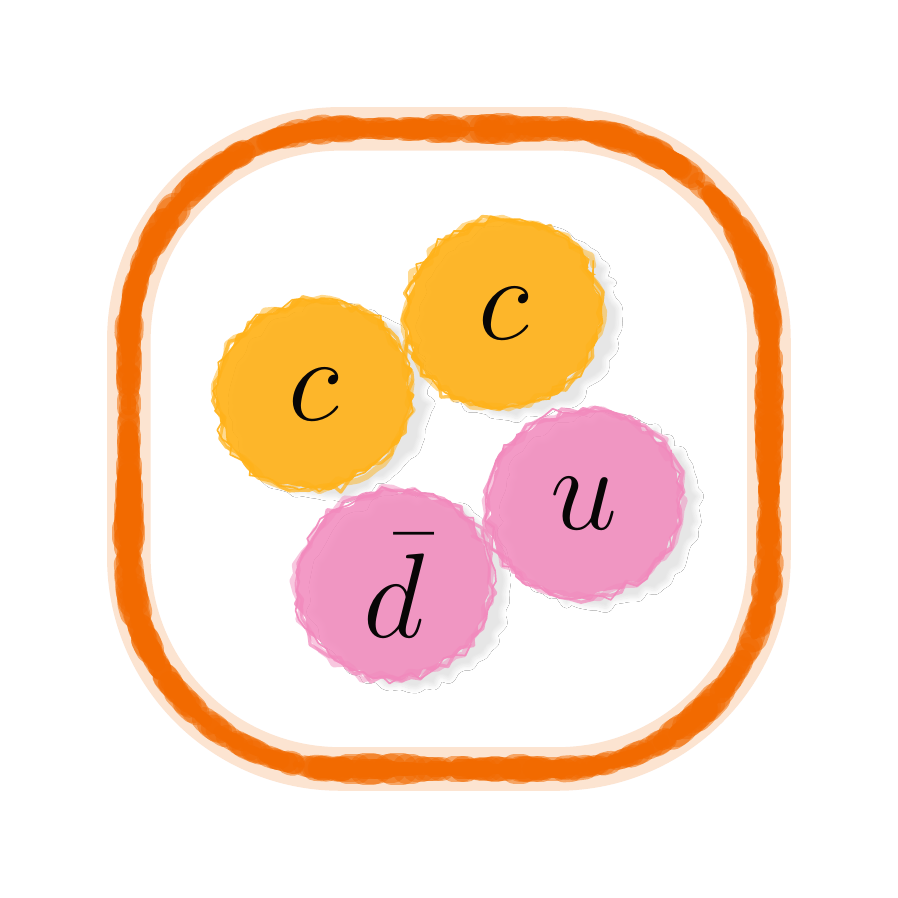}{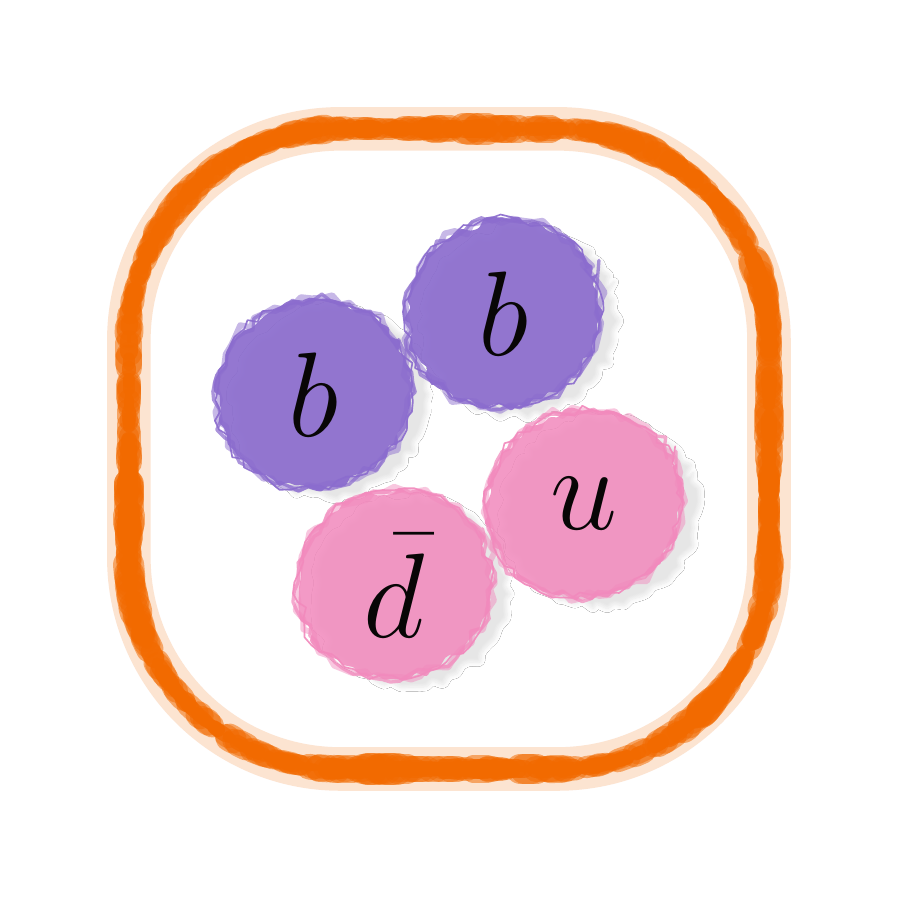}

The doubly-heavy tetraquark sector stands out as a textbook example of matter in four-quark configurations.
LHCb reported this extremely narrow and clean structure, named $T_{cc}^+$, from studies of the $D^0D^0\pi^+$ system produced inclusively in proton--proton collisions~\cite{LHCb:2022:tcc-discovery}.
The peak lies just below the $D^0D^{*+}$ threshold, with a binding of order a few hundred keV~\cite{LHCb:2022:tcc-study}, and has a width of about $50\,$keV.
The width measurement reflects the fact that the $D^0D^{*+}$ system is essentially elastic with two decay channels,
$D^0D^0\pi^+$ and $D^0D^0\gamma$. That constrains the width from above by the width of the $D^{*+}$ meson.
The simplicity of the system and the uniqueness of $T_{cc}^+$ have drawn considerable attention from phenomenological studies and from numerical lattice QCD computations.

\begin{figure}[h]
  \centering
  \raisebox{-0.5\height}{\includegraphics[width=0.49\textwidth]{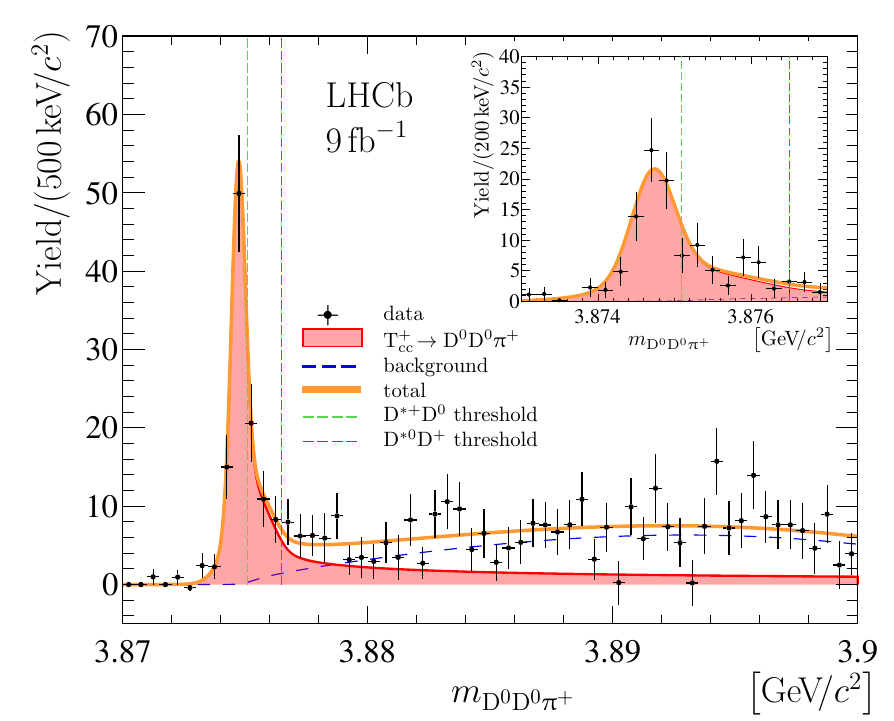}}
  \raisebox{-0.5\height}{\includegraphics[width=0.49\textwidth,trim=65mm 0 60mm 0,clip]{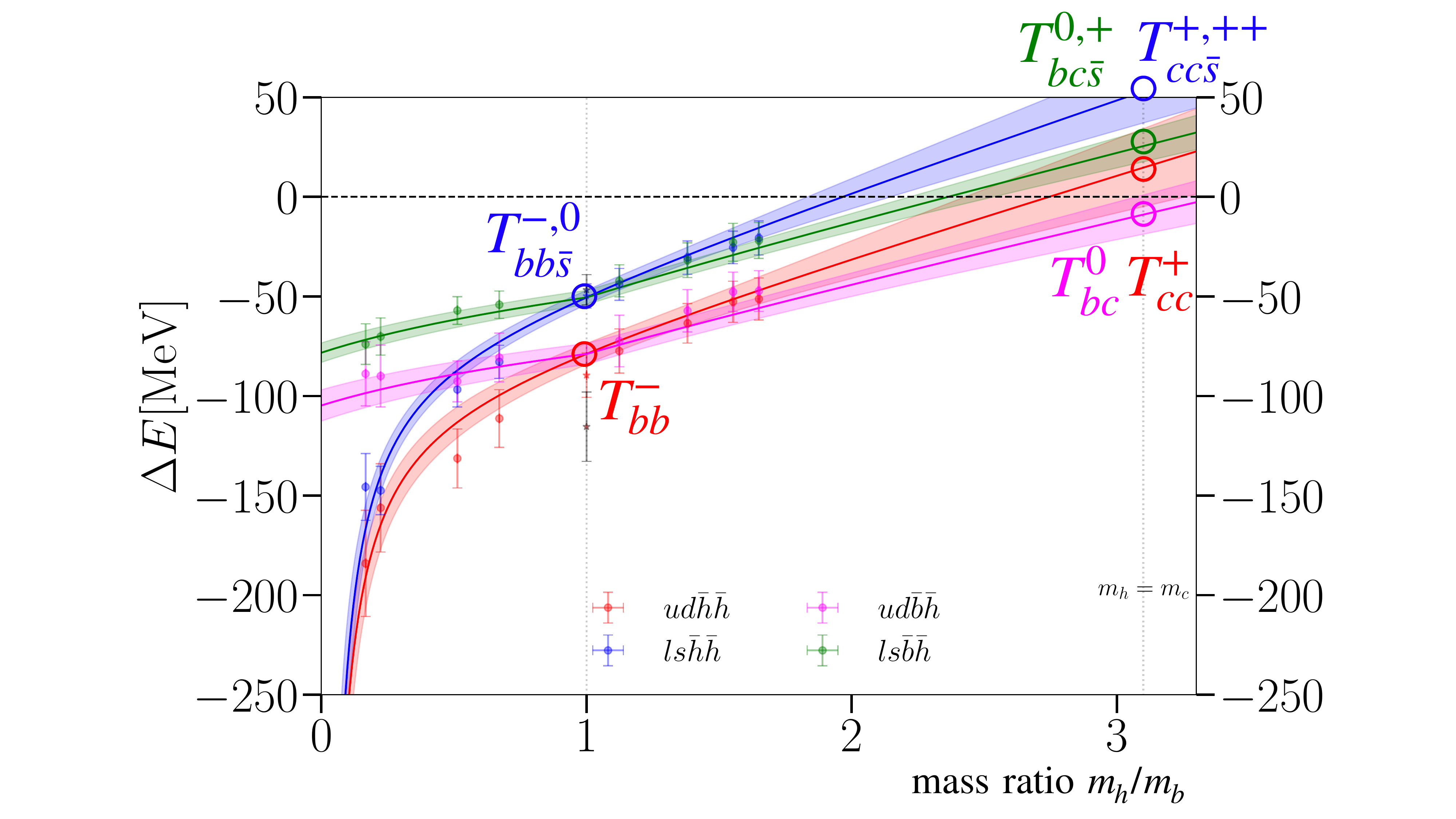}}
  \caption{Left: Mass spectrum of the $D^0D^0\pi^+$ system produced inclusively in proton--proton collisions, showing evidence for the doubly charmed $T_{cc}^+$ tetraquark.
  Right: Lattice QCD computation for the binding energy of doubly-heavy tetraquark families with quantum numbers $J^{P} = 1^{+}$.}
  \label{fig:Tcc_main}
\end{figure}

The existence of $T_{cc}^+$ also calls for its doubly-bottom partner, $T_{bb}^-$.
It is natural to expect that $T_{bb}^-$ should be more strongly bound
and stable with respect to the strong interaction.
A selected example of a quantitative estimate of the binding energy is given in Fig.~\ref{fig:Tcc_main} from recent lattice QCD calculations~\cite{Colquhoun:2024jzh}.
Because the heavy-flavor quarks are identical in the $cc$ and $bb$ cases, the heavy diquark ground state has quantum numbers $J^{P} = 0^{+}$,
suggesting $J^{P}=1^{+}$ for the ground state of the $T_{QQ}^+$ family.
The $T_{bc}^0$ state has both $J^{P}=0^{+}$ and $J^{P}=1^{+}$ components, with the former lower in mass.
While Fig.~\ref{fig:Tcc_main} suggests that the isovector $T_{bc}^0$ is bound, the theoretical uncertainty is large.
There is no consensus among theoretical predictions on the size of the binding for the scalar state relative to the $B^+D^-$ threshold.
Many other states within the SU(3) multiplet---$T_{bb\bar{s}}^{0,-}$ for double bottom,
$T_{bc\bar{s}}^{0,+}$ for bottom--charm,
and $T_{cc\bar{s}}^{+,++}$---are expected to exist.
They are unlikely to be narrow, however.
On the experimental side, challenges come from the smaller production cross section of $b$ quarks and from
the small branching fractions of $B$ mesons to any given exclusive decay mode.
Despite that, estimates suggest a chance of observing $T_{bc}^0$ using Run~3--4 datasets~\cite{Ivan.Tbc:2025xyz,Blusk.Tcc:2025xyz}. Prospects for $T_{bb}^-$ in exclusive decays are pessimistic, but there are ideas on how to improve statistics by using inclusive decays~\cite{Gershon:2018gda}.

\SubsectionQuarkTwo{Open-flavor tetraquarks}{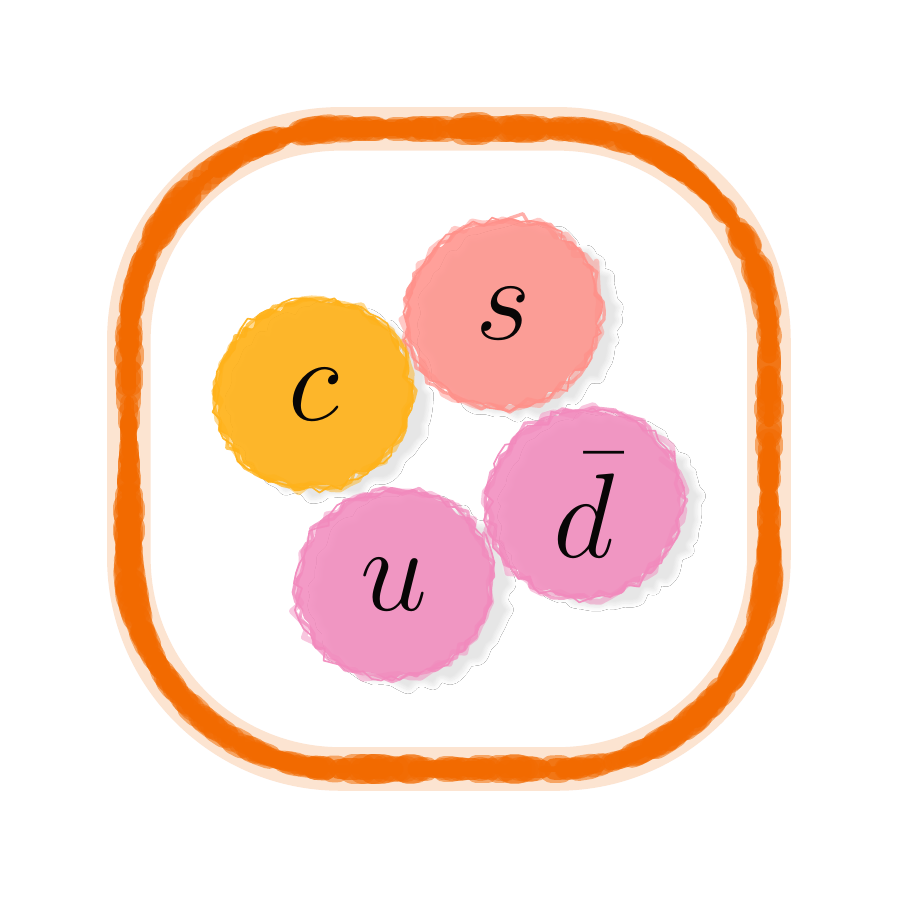}{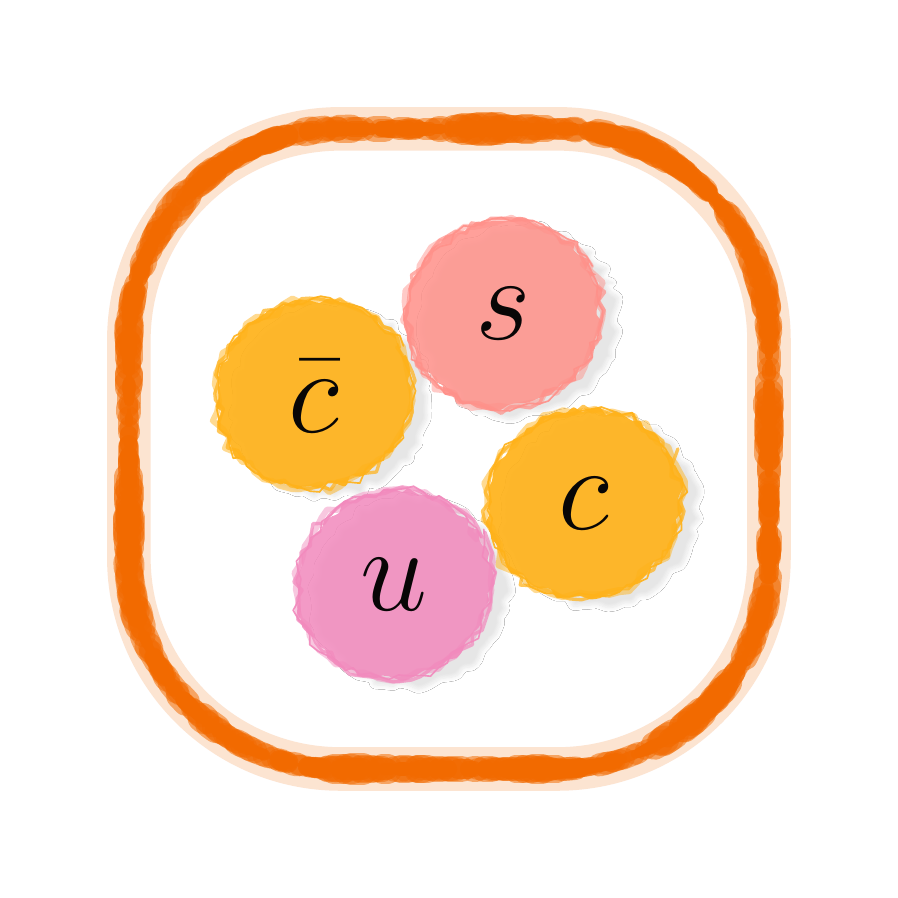}
\label{sec:B2DDh}

\begin{figure}[h]
  \centering
  \includegraphics[width=\textwidth,trim=55mm 0 60mm 0,clip]{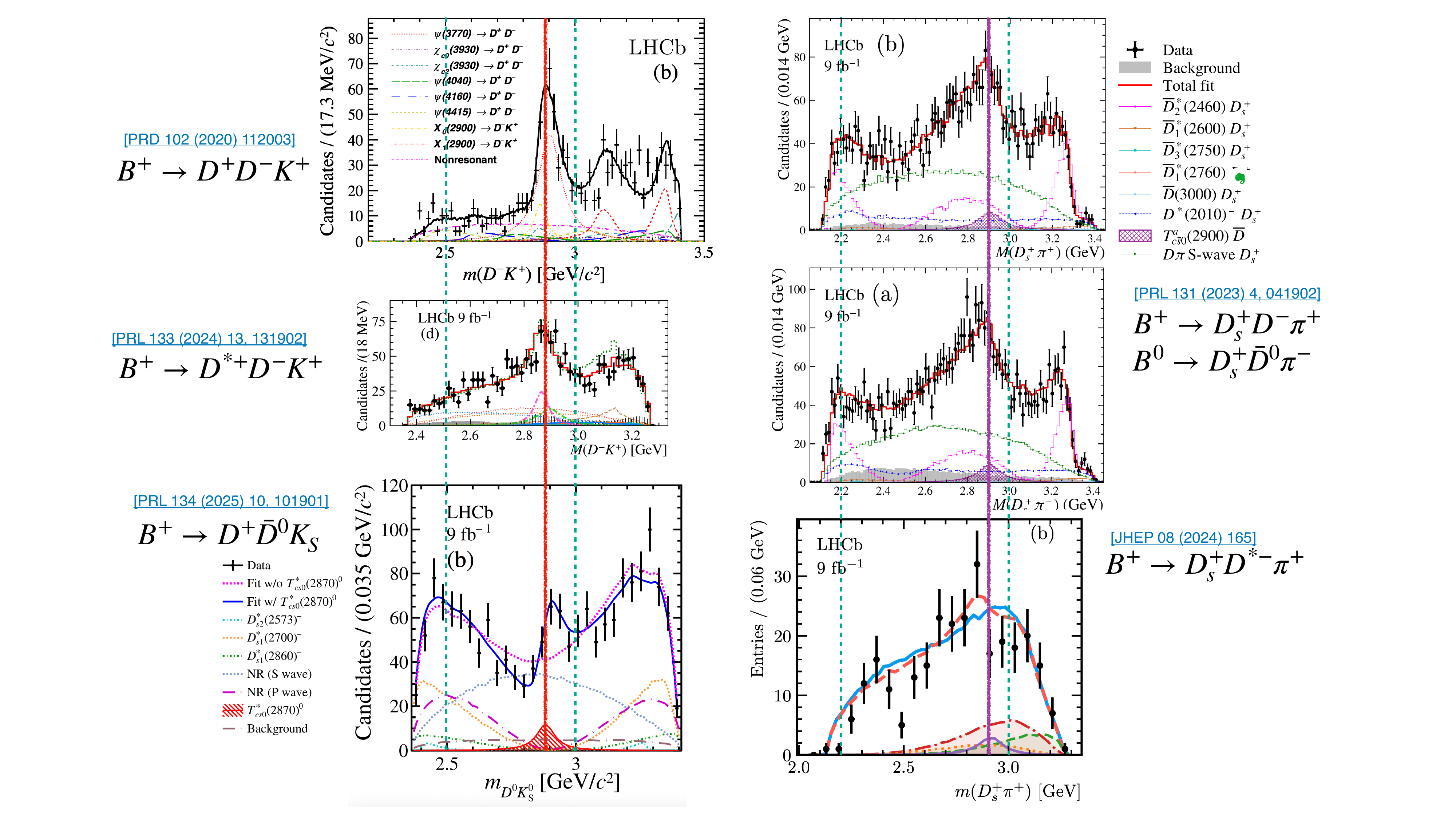}
  \caption{Overview of experimental results on charm-strange tetraquarks, $T_{c\bar s}$ and $T_{cs}$.}
  \label{fig:open-flavor-tetraquarks}
\end{figure}

Open-flavor tetraquarks---exotic states with one heavy quark (here $c$)---represent a very recent and not yet fully mapped sector of exotic spectroscopy.
It is experimentally active and will likely keep producing surprises even with current statistics.
There are two prominent exotic families: the $T_{cs}$ class and the $T_{c\bar s}$ class, shown in Fig.~\ref{fig:open-flavor-tetraquarks}.
The $T_{cs}$ candidates appear in $DK$-type systems, and the $T_{c\bar s}$ candidates in $D_s\pi$ combinations with exotic charge assignments.
In both cases we have clear experimental evidence of resonant enhancements,
but the detailed spectroscopy, including the number of states and their quantum numbers, requires further experimental input.

For the $T_{cs}$ sector,
the present picture is driven by a coherent chain of LHCb measurements in related decay modes, referred to as $B\to D\bar{D} h$ decays,
where $D$ stands for $D^+$, $D^0$, $D_s^+$, and $h$ stands for light (strange) hadron.
An amplitude analysis of $B^+ \to D^+D^-K^+$ first revealed a prominent peaking structure in the $D^-K^+$ channel near $2.9\,\mathrm{GeV}$~\cite{LHCb:2020pxc}. This was followed by the study of $B^+ \to D^{*\pm}D^{\mp}K^+$, where clear structures were again observed and the amplitude description favored two opposite-parity contributions~\cite{LHCb:2024vfz}.
More recently, an isospin-related signal was reported in $B^- \to D^-D^0K_S^0$~\cite{LHCb:2024xyx}.
Following the $B^+ \to D^+D^-K^+$ amplitude analysis, the enhancement is described by two overlapping resonances, $T_{cs0}(2900)^0$ and $T_{cs1}(2900)^-$.
Studies of other prominent decay channels, $B^+\to D^0\bar{D}^0K^+$ and $B^0\to D^-D^0K^+$, are available to date only from the older BaBar dataset~\cite{BaBar:2014jjr}, which reported no exotic signals at the time.

For the $T_{c\bar s}$ sector, the key observation is the appearance of peaking structures in $D_s^+\pi^+$ and $D_s^+\pi^-$
studied in the combined amplitude analysis of $B^+\to D_s\bar{D} h$ decays. The paper titled ``First Observation of a Doubly Charged Tetraquark and Its Neutral Partner''~\cite{LHCb:2022sfr} opened a new chapter in the study of exotics.
Because the $D_s^+$ has $I=0$, resonances in the $D_s^+\pi^+$ channel
provide direct evidence for nonconventional hadrons with at least four quarks.
A signal for the $T_{c\bar s}^+$ resonance in the subsequent amplitude analysis of $B^+ \to D_s^+D^{*-}\pi^+$ was not statistically significant~\cite{LHCb:2024vhs} due to limited statistics in that channel.

It is easy to get lost in many decays within the $B\to D\bar{D} h$ class.
I therefore attempted to systematize the overview in Table~\ref{tab:cf-modes}.
The Cabibbo-favored three-body decays of $\bar B_Q=(\bar b Q)$, with $Q=u,d,s$, are primarily driven by the quark-level transition $\bar b \to \bar c\, c\, \bar s$, supplemented by the creation of a light quark pair $q\bar q$ from the vacuum. The final quark content is therefore $\bar c$, $c$, $\bar s$, $Q$, $q$, and $\bar q$.
\begin{align*}
  (Q\,\bar{b}) \to (\bar{c}\,\Box) + (c\,\Box) + (\Box\,\Box).
\end{align*}
Here the empty boxes are to be filled with the quarks $Q,\bar{s},q,\bar{q}$.
These quarks can be grouped into three mesons in four distinct ways. We denote the corresponding classes as follows:
\begin{align*}
  \textrm{I}:   & (\bar c Q)(c\bar q)(\bar s q) &
  \textrm{II}:  & (\bar c Q)(c\bar s)(q\bar q)  &
  \textrm{III}: & (\bar c q)(c\bar q)(\bar s Q) &
  \textrm{IV}:  & (\bar c q)(c\bar s)(\bar q Q) & .
\end{align*}
In the first and second classes, the B-meson flavor tags the $\bar D$-type meson.
In the third and fourth classes, it tags the strange meson and the light meson, respectively.
The second and fourth classes contain $D_s^+$ mesons. Class~II contains a neutral $q\bar q$ meson.

\begin{figure}[t]
  \centering
\begin{tikzpicture}[scale=1.12]
\def\ResPieR{1.72}
\def\ResPieLabelR{1.224}
\def\ResPieRtag{2.064}
\fill[respieslice2] (0,0) -- (90:\ResPieR) arc (90:50:\ResPieR) -- cycle;
\node[font=\footnotesize, inner sep=1pt] at (70:\ResPieLabelR) {$D^{+,0}$};
\node[inner sep=0pt] at (70:\ResPieRtag) {\resmark{2}{1}};
\fill[respieslice2] (0,0) -- (50:\ResPieR) arc (50:10:\ResPieR) -- cycle;
\node[font=\footnotesize, inner sep=1pt] at (30:\ResPieLabelR) {$D_s^{+}$};
\node[inner sep=0pt] at (30:\ResPieRtag) {\resmark{2}{1}};
\fill[respieslice0] (0,0) -- (10:\ResPieR) arc (10:-30:\ResPieR) -- cycle;
\node[font=\footnotesize, inner sep=1pt] at (-10:\ResPieLabelR) {$\psi$};
\node[inner sep=0pt] at (-10:\ResPieRtag) {\resmark{0}{1}};
\fill[respieslice1] (0,0) -- (-30:\ResPieR) arc (-30:-70:\ResPieR) -- cycle;
\node[font=\footnotesize, inner sep=1pt] at (-50:\ResPieLabelR) {$T_{c\bar cs}^{0,-}$};
\node[inner sep=0pt] at (-50:\ResPieRtag) {\resmark{1}{0}};
\fill[respieslice3] (0,0) -- (-70:\ResPieR) arc (-70:-110:\ResPieR) -- cycle;
\node[font=\footnotesize, inner sep=1pt] at (-90:\ResPieLabelR) {$T_{cs}^{0}$};
\node[inner sep=0pt] at (-90:\ResPieRtag) {\resmark{3}{1}};
\fill[respieslice3] (0,0) -- (-110:\ResPieR) arc (-110:-150:\ResPieR) -- cycle;
\node[font=\footnotesize, inner sep=1pt] at (-130:\ResPieLabelR) {$T_{c\bar s}^{++,0}$};
\node[inner sep=0pt] at (-130:\ResPieRtag) {\resmark{3}{1}};
\fill[respieslice1] (0,0) -- (-150:\ResPieR) arc (-150:-190:\ResPieR) -- cycle;
\node[font=\footnotesize, inner sep=1pt] at (-170:\ResPieLabelR) {$T_{c\bar c}^{-}$};
\node[inner sep=0pt] at (-170:\ResPieRtag) {\resmark{1}{0}};
\fill[respieslice3] (0,0) -- (-190:\ResPieR) arc (-190:-230:\ResPieR) -- cycle;
\node[font=\footnotesize, inner sep=1pt] at (-210:\ResPieLabelR) {$T_{cs}^{+,-}$};
\node[inner sep=0pt] at (-210:\ResPieRtag) {\resmark{3}{0}};
\fill[respieslice4] (0,0) -- (-230:\ResPieR) arc (-230:-270:\ResPieR) -- cycle;
\node[font=\footnotesize, inner sep=1pt] at (-250:\ResPieLabelR) {$T_{c\bar s\bar s}^{++,+}$};
\node[inner sep=0pt] at (-250:\ResPieRtag) {\resmark{4}{0}};
\end{tikzpicture}
  \caption{
    Two-meson subsystems accessible within the $B\to D\bar{D} h$ analyses, as listed in Table~\ref{tab:cf-modes}, grouped by $u$/$d$ isospin multiplet.
    Crossed boxes indicate classes that are included in the experimental analyses.
  }
  \label{fig:cf-resonance-pie}
\end{figure}
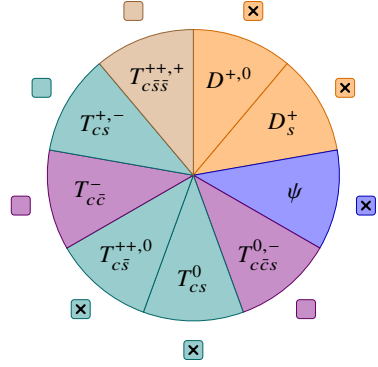

The number of distinct reactions amounts to 22 once the light quark--antiquark pairs $u\bar{u}$ and $d\bar{d}$ are identified, as listed in Table~\ref{tab:cf-modes}.
Of these, eight include charged light mesons in the final state, which is favorable experimentally; six include a neutral kaon;
the other five have $s\bar{s}$ mesons that can be accessed experimentally via $\phi$ mesons; and three have light $u\bar{u}$ or $d\bar{d}$ pairs.
The $D^0$, $D^+$, and $D_s^{+}$ mesons are reconstructed as $K^-\pi^+$, $K^-\pi^+\pi^+$, and $K^-K^+\pi^+$ decay modes, respectively.
\begin{table}[h]
  \centering
  \small
  \setlength{\tabcolsep}{4pt}%
  \renewcommand{\arraystretch}{1.06}%
  \caption{Cabibbo-favored three-body modes with clean charged-particle signatures. Marker hue encodes different classes of intermediate structures. Crossed boxes indicate classes that are included in the experimental analyses.}
  \label{tab:cf-modes}
  \begin{tabular}{@{}ll|cc|cc|cc|r@{}}
    Reaction & Class & \multicolumn{2}{c|}{$\bar D D$} & \multicolumn{2}{c|}{$D h$} & \multicolumn{2}{c|}{$h\bar D$} & Ref. \\
\hline
$B^{+} \to D^{-} D^{+} K^{+}$ & III & \resmark{0}{1} & \ressublabel{0}{1}{$\psi$} & \resmark{3}{0} & \ressublabel{3}{0}{$T_{c\bar s}^{++}$} & \resmark{3}{1} & \ressublabel{3}{1}{$T_{cs}^{0}$} & \cite{LHCb:2020bls,LHCb:2020pxc,LHCb:2024vfz} \\
$B^{+} \to \bar D^{0} D^{0} K^{+}$ & I, III & \resmark{0}{1} & \ressublabel{0}{1}{$\psi$} & \resmark{2}{1} & \ressublabel{2}{1}{$D_s^{+}$} & \resmark{3}{0} & \ressublabel{3}{0}{$T_{cs}^{-}$} & \cite{BaBar:2014jjr} \\
$B^{0} \to D^{-} D^{0} K^{+}$ & I & \resmark{1}{0} & \ressublabel{1}{0}{$T_{c\bar c}^{-}$} & \resmark{2}{1} & \ressublabel{2}{1}{$D_s^{+}$} & \resmark{3}{0} & \ressublabel{3}{0}{$T_{cs}^{0}$} & \cite{BaBar:2014jjr} \\
$B^{+} \to D_s^{-} D_s^{+} K^{+}$ & III, IV & \resmark{0}{1} & \ressublabel{0}{1}{$\psi$} & \resmark{4}{0} & \ressublabel{4}{0}{$T_{c\bar s\bar s}^{++}$} & \resmark{2}{1} & \ressublabel{2}{1}{$D^{0}$} & \cite{LHCb:2022aki} \\
$B^{+} \to D^{-} D_s^{+} \pi^{+}$ & IV & \resmark{1}{0} & \ressublabel{1}{0}{$T_{c\bar cs}^{0}$} & \resmark{3}{1} & \ressublabel{3}{1}{$T_{c\bar s}^{++}$} & \resmark{2}{1} & \ressublabel{2}{1}{$D^{0}$} & \cite{LHCb:2022sfr,LHCb:2024vhs} \\
$B^{0} \to \bar D^{0} D_s^{+} \pi^{-}$ & IV & \resmark{1}{0} & \ressublabel{1}{0}{$T_{c\bar cs}^{-}$} & \resmark{3}{1} & \ressublabel{3}{1}{$T_{c\bar s}^{0}$} & \resmark{2}{1} & \ressublabel{2}{1}{$D^{+}$} & \cite{LHCb:2022sfr} \\
$B_s^{0} \to D_s^{-} D^{0} K^{+}$ & I & \resmark{1}{0} & \ressublabel{1}{0}{$T_{c\bar cs}^{-}$} & \resmark{2}{0} & \ressublabel{2}{0}{$D_s^{+}$} & \resmark{2}{0} & \ressublabel{2}{0}{$D^{0}$} & --- \\
$B_s^{0} \to \bar D^{0} D_s^{+} K^{-}$ & IV & \resmark{1}{0} & \ressublabel{1}{0}{$T_{c\bar cs}^{-}$} & \resmark{2}{0} & \ressublabel{2}{0}{$D^{0}$} & \resmark{2}{0} & \ressublabel{2}{0}{$D_s^{+}$} & --- \\
\hline
$B^{+} \to \bar D^{0} D^{+} K^{0}$ & I & \resmark{1}{0} & \ressublabel{1}{0}{$T_{c\bar c}^{-}$} & \resmark{2}{1} & \ressublabel{2}{1}{$D_s^{+}$} & \resmark{3}{1} & \ressublabel{3}{1}{$T_{cs}^{0}$} & \cite{LHCb:2024xyx} \\
$B^{0} \to D^{-} D^{+} K^{0}$ & I, III & \resmark{0}{0} & \ressublabel{0}{0}{$\psi$} & \resmark{2}{0} & \ressublabel{2}{0}{$D_s^{+}$} & \resmark{3}{0} & \ressublabel{3}{0}{$T_{cs}^{+}$} & --- \\
$B_s^{0} \to D_s^{-} D^{+} K^{0}$ & I & \resmark{1}{0} & \ressublabel{1}{0}{$T_{c\bar cs}^{0}$} & \resmark{2}{0} & \ressublabel{2}{0}{$D_s^{+}$} & \resmark{2}{0} & \ressublabel{2}{0}{$D^{+}$} & --- \\
$B^{0} \to \bar D^{0} D^{0} K^{0}$ & III & \resmark{0}{0} & \ressublabel{0}{0}{$\psi$} & \resmark{3}{0} & \ressublabel{3}{0}{$T_{c\bar s}^{0}$} & \resmark{3}{0} & \ressublabel{3}{0}{$T_{cs}^{0}$} & --- \\
$B^{0} \to D_s^{-} D_s^{+} K^{0}$ & III, IV & \resmark{0}{0} & \ressublabel{0}{0}{$\psi$} & \resmark{4}{0} & \ressublabel{4}{0}{$T_{c\bar s\bar s}^{+}$} & \resmark{2}{0} & \ressublabel{2}{0}{$D^{+}$} & --- \\
$B_s^{0} \to D^{-} D_s^{+} \bar K^{0}$ & IV & \resmark{1}{0} & \ressublabel{1}{0}{$T_{c\bar cs}^{0}$} & \resmark{2}{0} & \ressublabel{2}{0}{$D^{+}$} & \resmark{2}{0} & \ressublabel{2}{0}{$D_s^{+}$} & --- \\
\hline

  \end{tabular}
\end{table}
The overview of the subchannel systems accessible within the $B\to D\bar{D} h$ class is shown in Fig.~\ref{fig:cf-resonance-pie}.

\subsection{Outlook}

The field of hadron spectroscopy has been growing rapidly in the last 15 years, driven by the new data from $pp$ and $e^+e^-$
machines. The largest number of contributions to the field of exotic hadrons has come from exclusive hadronic decays of $b$-hadrons.
Key measurements in double-heavy quark systems are not accessible in $b$-decays.
New candidate states in classes with multiple heavy quarks are instead observed in prompt production.

Hadronic phenomena appear progressively cleaner as the heavy-quark mass increases.
The multiquark sector likely follows the same pattern.
Tetraquark and pentaquark states in several distinct flavor classes can be related to one another through their quark-mass dependence.
The mechanisms governing the transition between bound-state and resonance descriptions remain to be clarified.
This question should be revisited as more precise data become available.
Many observed states are nonetheless found near thresholds, in patterns that are often consistent with hadronic molecules.

Most of the hadrons reviewed above fall into this pattern of either showing up as hadronic molecules ($P_{c\bar{c}}$, $P_{c\bar{c}s}$, $T_{cc}$), or having transitioned to the resonance region ($T_{cs}$, $T_{c\bar s}$, some $T_{c\bar c}$).
Within this picture, di-$J/\psi$ resonances are surprising and have not yet been connected to what is already known from the analogous sectors.

The prospects for spectroscopy at the LHC are bright, with the Run~3 data-taking campaign approaching completion in June~2026.
The results on the threefold LHCb data sample are only slowly starting to come in.
For purely muonic channels, CMS is gaining an advantage due to its higher luminosity, and the high-luminosity era around 2036 is on the horizon~\cite{Tomas:2025sky}.

For the Asian experiments, BESIII remains the main producer of unique data in the charmonium region from targeted energy scans, while approval of its successor, the Super Tau Charm Factory (STCF), is likely to remain a few years away~\cite{Ai:2025xop}.
In the coming years, the Belle~II detector will gather a data sample a few times larger than Belle's,
and will contribute substantially to heavy-flavor spectroscopy in channels not accessible to LHC experiments.
A major upgrade of the accelerator is planned during the long shutdown~3, starting in 2032~\cite{KEKBFactoryReview:2026}.

\section*{Acknowledgements}
\noindent
The author thanks colleagues in the LHCb collaboration for helpful discussions and comments on this contribution.

\bibliographystyle{unsrtnat}
\bibliography{master}

\end{document}